\documentclass[prb]{revtex4}
\usepackage{natbib}
\usepackage{psfig}
\usepackage{epsfig}
\usepackage{amssymb}
\usepackage{subfigure}
\usepackage{pstricks}
\usepackage{color}

\begin{document}
\renewcommand{\thesection}{\Roman{section}}
\renewcommand{\thesubsection}{\Alph{subsection}}
\input{epsf}
\pagestyle{plain}

 \title{An infinity of possible invariants for decaying homogeneous turbulence}
 \author{J. C. Vassilicos}
 \maketitle
\begin{center}
Turbulence, Mixing and Flow Control Group, Department of Aeronautics\\
Imperial College London, London, SW7 2BY, UK

Received (30 October 2010); Revised (28 December 2010)

\end{center}

%

\small{The von K\'arman-Howarth equation implies an infinity of
  invariants corresponding to an infinity of different asymptotic
  behaviours of the double and triple velocity correlation functions
  at infinite separations. Given an asymptotic behaviour at infinity
  for which the Birkhoff-Saffman invariant is not infinite, there are
  either none, or only one or only two finite invariants. If there are
  two, one of them is the Loitsyansky invariant and the decay of large
  eddies cannot be self-similar. We examine the consequences of this
  infinity of invariants on a particular family of exact solutions of
  the von K\'arman-Howarth equation.}


\newpage

\section{Introduction}
\label{threepone}

Results from recent laboratory experiments\cite{MV10} suggest that
classes of homogeneous turbulence decay exist which are at odds with
classical theory\cite{batch}. As the general theory of homogeneous
turbulence decay is based on invariants of the von K\'arman-Howarth
equation\cite{batch, batch2}, these recent experiments call for a
fresh study of what is true about these invariants. The present letter
provides such a study in the context of decaying homogeneous isotropic
turbulence. However, the assumption of isotropy could be dropped by
following, for example, the method of Nie \& Tanveer\cite{saleh}.

\section{Invariants of the von K\'arman-Howarth equation}
\label{threeptwo}

\noindent
Starting from the von K\'arman-Howarth equation for decaying
homogeneous isotropic turbulence\cite{batch}$^{,}$\cite{batch2}, we
show that it is possible to derive an infinite number of different
invariants corresponding to an infinite number of different conditions
at infinity. This equation is
\begin{equation}
{\partial \over \partial t} (u'^{2} f) = u'^{3} ({\partial k \over
  \partial r} + {4k\over r}) + 2\nu u'^{2}({\partial^{2} f \over
  \partial r^{2}} + {4\over r}{\partial f \over \partial r})
   \label{KH}
\end{equation}
where $u'= u'(t)$ is the r.m.s. of the turbulent fluctuating velocity
component $u$, $u'^{2} f(r,t) \equiv <u(x,t) u(x+r,t)>$ and $u'^{3}
k(r,t) \equiv <u^{2}(x,t) u(x+r,t)>$, the brackets signifying an
average over realisations or over the spatial coordinate $x$ which is
defined on the same axis as the velocity component $u$. Note that
$r\ge 0$, that $f(0,t)=1$ and that reflection invariance implies
$k(0,t)=0$. It is natural to assume that all derivatives of $f$ and
$k$ with respect to $r$ are not infinite at $r=0$.


Given suitable conditions at infinity, equation (\ref{KH}) can be used
to calculate the rate of change of $u'^{2} \int_{0}^{+\infty} r^{m}
{\partial^{n} f \over \partial r^{n}} dr$ for an infinite range of values
of $m$ and $n$. Repeated integrations by parts yield
 $$
{d\over dt} \left[ u'^{2} \int_{0}^{+\infty} r^{m} {\partial^{n} f
    \over \partial r^{n}} dr \right] =
$$ 
\begin{equation}
(-1)^{n}u'^{3}\int_{0}^{+\infty} dr r^{m-n-1} k(r)[4T_{m-n+1}^{m} -
    T_{m-n}^{m}] + (-1)^{n}2\nu u'^{2} \int_{0}^{+\infty} dr r^{m-n-2}
  f(r)[T_{m-n-1}^{m} - 4 (m-n-1)T_{m-n+1}^{m}]
\label{mn}
\end{equation}
where $T_{m+p}^{m} =1$ and $T_{m-p}^{m} = m(m-1)...(m-p)$
if $p$ is a positive integer, and $T_{m}^{m} =m$ (note that $m$ does
not need to be an integer).  These integrations by parts yield the
right-hand side of (\ref{mn}) provided that 
$m > n+1$, $n\ge 0$, $\lim_{r\to \infty} (r^{m-n} k)=0$ and
$\lim_{r\to \infty} (r^{m-n-1} f)=0$. The integral $\int_{0}^{+\infty}
r^{m} {\partial^{n} f \over \partial r^{n}}dr$ is finite if
$\lim_{r\to \infty} (r^{m-n+1} f)=0$. We make the assumption that
$f(r)$ and $k(r)$ do not oscillate at infinity.

Noting that $T_{m-n-1}^{m} - 4 (m-n-1)T_{m-n+1}^{m} = (1+n-m)
[4T_{m-n+1}^{m} - T_{m-n}^{m}]$ for all $n\ge 0$, (\ref{mn}) simplifies to
 $$
{d\over dt} \left[ u'^{2} \int_{0}^{+\infty} r^{m} {\partial^{n} f
    \over \partial r^{n}} dr \right] =
$$ 
\begin{equation}
(-1)^{n} (4T_{m-n+1}^{m} - T_{m-n}^{m}) \left[ u'^{3}\int_{0}^{+\infty} dr
    r^{m-n-1} k(r) + 2(1+n-m) \nu u'^{2} \int_{0}^{+\infty} dr
    r^{m-n-2} f(r) \right] . 
\label{mn2}
\end{equation}
By considering linear combinations of pairs of integrals
$\int_{0}^{+\infty} r^{m} {\partial^{n} f \over \partial r^{n}} dr$
and $\int_{0}^{+\infty} r^{m'} {\partial^{n'} f \over \partial r^{n'}}
dr$ for which $m-n=m'-n'\equiv M$, we can form an infinite number of
invariants. There are two cases. One where $M \equiv m-n=4$, in which
case there is no need to consider such linear combinations because
$[4T_{m-n+1}^{m} - T_{m-n}^{m}] = [4T_{5}^{m} - T_{4}^{m}]=0$ for any
integer $m \ge 4$. This case immediately yields
\begin{equation}
{d\over dt} \left[ u'^{2} \int_{0}^{+\infty} r^{4+n} {\partial^{n} f
    \over \partial r^{n}} dr \right] = 0
\label{Loitn}
\end{equation}
for any integer $n\ge 0$ under the conditions $\lim_{r\to \infty}
(r^{4} k)=0$ and $\lim_{r\to \infty} (r^{5} f)=0$. These conditions
ensure that the quantity $ u'^{2} \int_{0}^{+\infty} r^{4+n}
{\partial^{n} f \over \partial r^{n}} dr $ is both finite and
  independent of time for any integer $n\ge 0$. When $n=0$, this
  quantity is the well-known Loitsyansky
  invariant\cite{loit}$^{,}$\cite{batch2}. Integrations by parts show
  that this quantity is proportional to the Loitsyansky invariant for
  any $n\ge 0$ because $\lim_{r\to \infty} (r^{5} f)=0$ and $f(r)$ is
  assumed not to oscillate at infinity.

The second case is for $M\not = 4$. In this case the following linear
combinations of integrals $\int_{0}^{+\infty} r^{m} {\partial^{n} f
  \over \partial r^{n}} dr$ and $\int_{0}^{+\infty} r^{m'}
{\partial^{n'} f \over \partial r^{n'}} dr$ are invariant:
\begin{equation}
I_{Mnn'} \equiv u'^{2}\int_{0}^{+\infty} r^{M+n'} {\partial^{n'} f(r)\over \partial r^{n'}} dr + C_{Mnn'} u'^{2}
\int_{0}^{+\infty} r^{M+n} {\partial^{n} f(r)\over \partial r^{n}} dr
\label{Inn'}
\end{equation}
where $m-n=m'-n'\equiv M \not = 4$, $n$ and $n'$ are non-negative
integers such that $n \not = n'$ and $C_{Mnn'} = -(-1)^{n'-n}
[4T_{M+1}^{M+n'} - T_{M}^{M+n'}]/[4T_{M+1}^{M+n} - T_{M}^{M+n}]$. From
(\ref{mn2}),
\begin{equation}
{d\over dt} I_{Mnn'} =0
\label{0}
\end{equation}
under the conditions that $M>1$, $\lim_{r\to \infty} (r^{M} k)=0$ and
$\lim_{r\to \infty} (r^{M-1} f)=0$ and that $I_{Mnn'}$ is
well-defined. Hence, the von K\'arman-Howarth equation admits an
infinity of possible finite integral invariants depending on
conditions at infinity.


Whilst $M$ does not have to be an integer, the smallest integer value
of $M$ for which such invariants exist is $M=2$. The particular choice
$M=2$, $n'=0$ and $n=1$ recovers the Birkhoff-Saffman
invariant\cite{birk}$^{,}$\cite{saff}
\begin{equation}
3 I_{210} = u'^{2}\int_{0}^{+\infty} \left[3 r^{2} f(r) + r^{3} {\partial
  f(r)\over \partial r }\right] dr.
\label{BS}
\end{equation}
The use of a single integral in this expression instead of the two
integrals in equation (\ref{Inn'}) is significant because $3 r^{2}
f + r^{3} {\partial f \over \partial r } = {\partial \over
  \partial r } (r^{3} f)$ leads to
\begin{equation}
3 I_{210} = u'^{2} \lim_{r\to \infty} (r^{3} f), 
\label{3}
\end{equation}
showing that $I_{210} =0$ if $\lim_{r\to \infty} (r^{3} f) =0$, but also
that $I_{210}$ takes a finite value if defined as in (\ref{BS}) rather
than (\ref{Inn'}) and if $\lim_{r\to \infty} (r^{3} f)$ is finite.

The Birkhoff-Saffman invariant (\ref{BS}) can be generalised into an
infinite series of invariants in two steps. Firstly, for any $n\ge 1$,
define
\begin{equation}
I_{2n0} = u'^{2}\int_{0}^{+\infty} \left[r^{2} f(r) + C_{2n0} r^{2+n}
  {\partial^{n} f(r)\over \partial r^{n}}\right] dr
\label{BSinf}
\end{equation}
for which the following iterative relation holds:
\begin{equation}
I_{2(n+1)0} = I_{2n0} + C_{2(n+1)0} u'^{2} \lim_{r\to \infty} (r^{3+n}
{\partial^{n} f\over \partial r^{n}}).
\label{BSiter}
\end{equation}
Hence, if $f(r,t)\approx a_{3}(t) (L(t)/r)^{3}$ (where $L(t)$ is a
length-scale and $a_{3} L^{3}\not \equiv 0$) to leading order when
$r\to\infty$, then the generalised Birkhoff-Saffman invariants
$I_{2n0}$ are finite and their time-independence implies the
time-independence of $a_{3} L^{3} u'^{2}$ (and vice versa). As a
second step, define
\begin{equation}
I_{Mnn'} \equiv u'^{2} \int_{0}^{+\infty} \left[r^{M+n'}
  {\partial^{n'} f(r)\over \partial r^{n'}} + C_{Mnn'} r^{M+n}
  {\partial^{n} f(r)\over \partial r^{n}} \right] dr
\label{BSinf2}
\end{equation}
for any $M>1$ and any $n \not = n'$. Noting that $C_{Mn(n'+1)}=
-(M+n'+1)C_{Mnn'}$, one can derive the iterative relation
\begin{equation}
I_{Mn(n'+1)} = -(M+n'+1)I_{Mnn'} + u'^{2} \lim_{r\to \infty} (r^{M+n'+1}
{\partial^{n'} f\over \partial r^{n'}}).
\label{BSiter2}
\end{equation} 
Under the same assumption that $f(r,t)\approx a_{3}(t) (L(t)/r)^{3}$
to leading order when $r\to\infty$, (\ref{BSinf2}) and (\ref{BSiter2})
applied to $M=2$ can now be used to show that all generalised
Birkhoff-Saffman invariants $I_{2nn'}$ are finite and their
time-independence is equivalent to the time-independence of $a_{3}
L^{3} u'^{2}$.

Hence, our generalised Birkhoff-Saffman invariants lead to a
conclusion previously reached by Birkhoff\cite{birk} and
Saffman\cite{saff} on the basis of the constancy of (\ref{BS})
alone. Namely, if $f(r)\approx a_{3} (L/r)^{3}$ as $r\to\infty$, and
if $\lim_{r\to \infty} (r^{2} k)=0$, then
\begin{equation}
{d\over dt} (a_{3} L^{3} u'^{2})=0.
\label{BSa}
\end{equation}

As for any $M>1$ but different from 2 and 4, the case $M=3$
corresponds to a new set of integral invariants. Similarly to the
$M=2$ case, we rewrite the invariants $I_{3n0}$ using only one
integral, i.e.
\begin{equation}
I_{3n0} = u'^{2}\int_{0}^{+\infty} \left[r^{3} f(r) + C_{3n0} r^{3+n}
  {\partial^{n} f(r)\over \partial r^{n}}\right] dr
\label{JCV}
\end{equation}
for $n\ge 1$, and we note that
\begin{equation}
4 I_{310} = u'^{2} \lim_{r\to \infty} (r^{4} f)
\label{4}
\end{equation}
and that 
\begin{equation}
I_{3(n+1)0} = I_{3n0} + C_{3(n+1)0} u'^{2} \lim_{r\to \infty} (r^{4+n}
{\partial^{n} f\over \partial r^{n}}). 
\label{JCViter}
\end{equation}

The condition $\lim_{r\to \infty}(r^{4} f)=0$ under which we
established the constancy of $I_{3n0}$ implies $I_{310}=0$. However,
if $I_{3n0}$ is defined as in (\ref{JCV}) rather than (\ref{Inn'}),
then it is permitted to relax this condition and assume instead that
$f(r,t)\approx a_{4}(t) (L(t)/r)^{4}$ (where $L(t)$ is a length-scale
and $a_{4} L^{4}\not \equiv 0$) to leading order when $r\to\infty$. In
this case, and without forgetting the accompanying condition
$\lim_{r\to \infty}(r^{3} k)=0$, $I_{3n0}$ is finite for all $n\ge 1$,
and its invariance in time leads to 
\begin{equation}
{d\over dt} (a_{4} L^{4} u'^{2})=0. 
\label{JCVa}
\end{equation}
An effectively identical argument to the one given above for
$I_{2nn'}$ shows that all integral invariants $I_{3nn'}$ are in fact
finite and time-independent under the conditions that $f(r,t)\approx
a_{4}(t) (L(t)/r)^{4}$ to leading order when $r\to\infty$ and
$\lim_{r\to \infty}(r^{3} k)=0$. Their time independence is also
equivalent to (\ref{JCVa}).


The cases $M>4$ are similar to the cases $M=2$ and $M=3$. In general,
for any $M > 1$ such that $M \not = 4$, we have
\begin{equation}
(M+1) I_{M10} = u'^{2} \lim_{r\to \infty} (r^{M+1} f)
\label{M}
\end{equation}
and
\begin{equation}
I_{M(n+1)0} = I_{Mn0} + C_{M(n+1)0} u'^{2} \lim_{r\to \infty} (r^{M+1+n}
{\partial^{n} f\over \partial r^{n}}). 
\label{iter}
\end{equation}
For simplicity, we focus on $I_{Mn0}$ because the argument based on
(\ref{BSinf2}) and (\ref{BSiter2}) which we gave for $M=2$ can be
applied here to show that what holds for $I_{Mn0}$ also holds for
$I_{Mnn'}$.  A re-definition of $I_{Mn0}$ in terms of a single
integral (\ref{BSinf2}) instead of (\ref{Inn'}) allows the possibility
for non-zero invariants of order $M$. Specifically, with such a
re-definition, it is possible to assume $f(r,t)\approx a_{M+1}(t)
(L(t)/r)^{M+1}$ (where $L(t)$ is a length-scale and $a_{M+1}
L^{M+1}\not \equiv 0$) to leading order when $r\to\infty$. In terms of
the energy spectrum $E(\kappa )$ in Fourier space, this assumption
takes the form $E( \kappa )\sim a_{M+1} u'^{2} L (\kappa L)^{M}$ in
the limit $\kappa \to 0$ when $M>1$ (because\cite{batch} $E(\kappa) =
{u'^{2}\over \pi} \int_{0}^{\infty} dr (3f(r) + r \partial f/\partial
r) \kappa r \sin \kappa r$). Under this assumption and the
accompanying condition $\lim_{r\to \infty}(r^{M} k)=0$, $I_{Mn0}$ is
finite for all $n\ge 1$ and its invariance leads to
\begin{equation}
{d\over dt} (a_{M+1} L^{M+1} u'^{2})=0. 
\label{a}
\end{equation}
This proves a more precise version of the principle of permanence of
large eddies given in p. 113 of the 1995 book by Frisch\cite{frisch}.
Note that (\ref{a}) has already been obtained 
by Rotta\cite{rotta} and Lundgren\cite{lundgren} by direct inspection
but without noticing the integral invariants (\ref{BSinf2}) and
therefore without the resulting systematic approach given here.

We stress that $M$ does not need to be an integer for equations
(\ref{0}), (\ref{M}), (\ref{iter}) and (\ref{a}) to hold. However, as
Rotta\cite{rotta} remarked, $E(k)$ results from an integral over a
spherical shell in wavenumber space\cite{batch} so that any $M<2$
would imply that the spectral tensor\cite{batch} (that is the Fourier
transform of the velocity correlation tensor $R_{ij} \equiv
<u_{i}({\bf x}) u_{j}({\bf x} + {\bf r})>$) diverges as $k\to 0$. We
therefore limit the remainder of this letter to $M\ge 2$. There is no
a priori upper limit to $M$ as the results of this section are valid
for any $M>1$.


\section{Consequences of these invariants}
\label{threepthree}

It is clear that we have an infinity of possible invariants depending
on the asymptotic behaviours of $f(r,t)$ and $k(r,t)$ at
infinity. Some of these invariants can also be expressed in terms of
the velocity correlation tensor $R_{ij}$, 
specifically in terms of its trace $R_{ii}$ which is a function of
only $r=\vert {\bf r} \vert$ because of homogeneity and isotropy. In
Batchelor's book on turbulence\cite{batch} one can find the identity
$R_{ii}(r)=u'^{2}(3f+r{\partial f\over \partial r})$ for homogeneous
isotropic turbulence. Using this identity, one obtains
\begin{equation}
\int r^{M-2}R_{ii} d{\bf r} = 4\pi \int_{0}^{\infty} r^{M} R_{ii}(r)
dr = 4\pi (M-2) u'^{2}\int_{0}^{\infty} r^{M} f(r) dr + 4\pi u'^{2}
\lim_{r\to \infty} (r^{M+1} f).
\label{R}
\end{equation}
for any $M\ge 2$. 

As noted by Birkhoff\cite{birk} and Saffman\cite{birk}, this integral
equals $4\pi u'^{2} \lim_{r\to \infty} (r^{3} f)$ when $M=2$ and is
finite if this limit is also finite. If this limit vanishes, then so
does $\int R_{ii} d{\bf r}$, but in both cases $\int R_{ii} d{\bf r}$
is an invariant.

For any $M>2$, $\int r^{M-2}R_{ii} d{\bf r}$ diverges in the case
where $\lim_{r\to \infty} (r^{M+1} f)$ is finite but equals $4\pi
(M-2) u'^{2}\int_{0}^{\infty} r^{M} f(r) dr$ in the case where
$\lim_{r\to \infty} (r^{M+1} f)=0$. Hence, with the exception of $M=2$
and $M=4$, $\int r^{M-2}R_{ii} d{\bf r}$ is not in general invariant,
even though there are invariants $I_{Mnn'}$ for every value of $M\ge
2$. (The case $M=4$ corresponds to $\int r^{2}R_{ii} d{\bf r}$ which
is, in fact, the Loitsyansky invariant in a different guise.)


We now show that, for conditions at infinity which are such that the
Birkhoff-Saffman invariant is not infinite, either none or only one or
only two invariants are finite.  Assuming that there exists a number
$M_{f}\ge 2$ for which $\lim_{r\to \infty} (r^{M_{f}+1} f)=
a_{M_{f}+1} L^{M_{f}+1} \not \equiv 0$ and a number $M_g$ for which
$\lim_{r\to \infty} (r^{M} k)=0$ for any $M$ in the interval $2\le M <
M_{g}$ but $\lim_{r\to \infty} (r^{M} k) \not = 0$ for any $M\ge
M_{g}$, then the following five possibilities present themselves for
$I_{Mnn'}$ redefined in terms of a single integral (\ref{BSinf2})
instead of (\ref{Inn'}).

\noindent
(i) $M_{g} < M_{f}$ and $M_{g}<4$, in which case all invariants
$I_{Mnn'} =0$ for $M < M_{g}$ and all $I_{Mnn'}$ for $M\ge M_{g}$ are not
invariant. 

\noindent
(ii) $M_{g} < M_{f}$ and $M_{g} \ge 4$, in which case all invariants
$I_{Mnn'} =0$ for $M < M_{g}$ except the Loitsyansky invariant which
is the single non-vanishing invariant, and all $I_{Mn}$ for $M\ge
M_{g}$ are not invariant. In this case $u'^{2}\int_{0}^{+\infty} r^{4}
f(r) dr$ is the only non-vanishing invariant.

\noindent
(iii) $4> M_{g} \ge M_{f}$, in which case all invariants $I_{Mnn'} =0$
for $M\le M_{f}$ but invariant $I_{M_{f} nn'} \not =0$ and all integrals
$I_{Mnn'}$ with $M>M_{f}$ diverge. 


\noindent
(iv) $M_{g}\ge 4 > M_{f}$ in which case all integrals $I_{Mnn'}$ for
which $M < M_{f}$ are invariant but vanish and $I_{M_{f} nn'} \not =0$
and is invariant.

\noindent
In cases (iii) and (iv), $a_{M_{f}+1} L^{M_{f}+1} u'^{2}$ is the only
non-vanishing invariant and $M_{f} < 4$.


\noindent
(v) $M_{g}\ge M_{f} \ge 4$ in which case there are only two
non-vanishing invariants when $M_{f}>4$, the Loitsyansky invariant and
$I_{M_{f} nn'}$, i.e. $u'^{2}\int_{0}^{+\infty} r^{4} f(r) dr$ and
$a_{M_{f}+1} L^{M_{f}+1} u'^{2}$. When $M_{f}=4$,
$u'^{2}\int_{0}^{+\infty} r^{4} f(r) dr$ is the sole non-vanishing
invariant.

All in all, depending on conditions at infinity, either no finite
invariants exist, or, if such exists, then either $a_{M_{f}+1}
L^{M_{f}+1} u'^{2}$ is the sole finite invariant with $M_{f}<4$, or
$u'^{2}\int_{0}^{+\infty} r^{4} f(r) dr$ is the sole finite invariant,
or $u'^{2}\int_{0}^{+\infty} r^{4} f(r) dr$ and $a_{M_{f}+1}
L^{M_{f}+1} u'^{2}$ (with $M_{f}>4$) are the only two finite
invariants.

We close this letter by testing this conclusion on George-type
self-preserving
solutions\cite{geor}$^{,}$\cite{geor2}$^{,}$\cite{MV10} of (\ref{KH})
because of the recent claim that it might be possible to engineer 
self-preserving decaying homogeneous isotropic turbulence in the wind
tunnel\cite{MV10}. These solutions are of the form $f(r,t) =
f[r/l(t)]$ and $k(r,t) = b(\nu, u'_{0}, l_{0}, t-t_{0})\kappa
[r/l(t)]$ where $u'_{0} \equiv u'(t_{0})$ and $l_{0} \equiv
l(t_{0})$. Introducing these forms into (\ref{KH}), one obtains the
solvability conditions ${d\over dt} u'^{2} = - 2\alpha \nu
u'^{2}/l^{2}$, ${d\over dt} l^{2} = c\nu $ and $b=\beta \nu/(u'l)$
where $\alpha >0 $, $c >0$ and $\beta$ are numerical constants. It
follows that
\begin{equation}
u'^{2}(t)= u'^{2}_{0} \left[ 1 + {c\nu\over
    l_{0}^{2}}(t-t_{0}) \right]^{-2\alpha/c}
\label{George1}
\end{equation}
and
\begin{equation}
l^{2}(t)= l_{0}^{2} +c\nu (t-t_{0}). 
\label{George2}
\end{equation}

If the conditions at infinity are such that no finite invariant
exists, then no obvious constraint can be imposed on the exponent
$2\alpha/c$ and the rate of turbulence decay. However, in the case
where the sole finite invariant is the Loitsyansky integral, then
$2\alpha/c = 5/2$. In the case where the sole finite invariant is
$a_{M_{f}+1} L^{M_{f}+1} u'^{2}$ with $2\le M_{f}<4$, then we can take
$L(t)=l(t)$ and the self-preserving form of $f$ implies that
$a_{M_{f}+1}$ must be constant in time. We therefore get $2\alpha/c =
(M_{f}+1)/2 $ which lies between $3/2$ and $5/2$.

Finally, when the conditions at infinity are such that
$u'^{2}\int_{0}^{+\infty} r^{4} f(r) dr$ and $a_{M_{f}+1} L^{M_{f}+1}
u'^{2}$ (where $M_{f}>4$) are both finite and invariant, then no
George-type self-preserving solution of (\ref{KH}) is allowed because
of the time-independence of $a_{M_{f}+1}$ implied by such
solutions. Noting that the contribution to $\int_{0}^{+\infty} r^{4}
f(r) dr$ coming from small values of $r$ is negligible, this
conclusion is valid more broadly for any form of $f(r)$ which is
permissible by (\ref{KH}) and the incompressible Navier-Stokes
equations and which conforms with self-similar decay of large
eddies\cite{frisch}, i.e.  for which $f(r,t) \approx f[r/l(t)]$ if $r$
is large enough and $a_{M_{f}+1}$ is time-independent as a
result. Hence, if $f(r)$ decays faster than $r^{-5}$ as $r\to \infty$
(i.e. $E(k)$ drops faster than $k^{4}$ as $k\to 0$), and if the
asympotic behaviour of the triple velocity correlation function is
such that two finite invariants exist at once (case (v) above), then
the decay of the large eddies cannot be self-similar.


\section{Conclusions}
\label{threepfour}

A summary of main conclusions is in the abstract. The nature of
turbulence decay depends critically on the asymptotic behaviour of the
double and triple velocity correlation functions at infinite
separations. There are four cases depending on whether $M_{f}/M_{g}$
is larger or smaller than 1 and whether $min (M_{f}, M_{g})$ is larger
or smaller than 4.

When $M_{f}/M_{g}$ is larger than 1 and $min (M_{f}, M_{g})$ is
smaller than 4 there are no finite invariants. When $M_{f}/M_{g}$ is
larger than 1 but $min (M_{f}, M_{g})$ is larger than 4 there is only
one finite invariant and this is the Loitsyansky invariant. When
$M_{f}/M_{g}$ is smaller than 1, there is either one or two finite
invariants dependending on whether $min (M_{f}, M_{g})$ is smaller or
larger than 4. In both cases $a_{M_{f}+1}L^{M_{f}+1} u'^{2}$ is finite
and invariant but when $min (M_{f}, M_{g})$ is larger than 4,
Loitsyansky's $u'^{2}\int_{0}^{+\infty} r^{4} f(r) dr$ is a finite
invariant too.

Self-preserving turbulence decays in accordance with (\ref{George1})
and (\ref{George2}) and the infinity of possible invariants permitted
by (\ref{KH}) cannot determine the exponent in (\ref{George1}) without
prior knowledge of correlations between points in the turbulence which
are extremely far apart. In fact, these correlations can even be such
that no conclusion whatsoever can be made on the value of the exponent
in (\ref{George1}), and the relatively high values reported for this
exponent in some wind tunnel experiments\cite{MV10} cannot be ruled
out theoretically without prior knowledge of these correlations. The
self-preserving decay which seems to have been observed in some
instances of fractal-generaged homogeneous turbulence\cite{MV10}
suggests that $M_f$ and $M_g$ cannot be such that $4 < M_{f} < M_{g}$
in such instances of turbulence if it is isotropic. Research with many
fundamentally different ways of generating turbulence\cite{MV10} needs
to be carried out so as to gain some understanding of what determines
conditions at infinity and whether they are all physically possible.


{\bf Acknowledgements:} I am grateful to B\'ereng\`ere Dubrulle,
Susumu Goto, Jonathan Gustafsson, John Hinch, Tom Lundgren, David
Thomson and Pedro Valente for kindly reading the first version of this
letter and offering valuable comments and suggestions which have
helped me improve it very significantly.

\makeatletter
\renewcommand{\@biblabel}[1]{\textsuperscript{#1}}
\makeatother

   \renewcommand{\section}[2]{}

\end{document}